\documentclass[english,floatfix,aps,prb,superscriptaddress,onecolumn]{revtex4-1}
\usepackage[T1]{fontenc}
\usepackage[utf8]{luainputenc}
\usepackage{amsmath}
\usepackage{amssymb}
\usepackage{stackrel}
\usepackage{graphicx}
\usepackage{esint}

\makeatletter

\@ifundefined{textcolor}{}
{%
 \definecolor{BLACK}{gray}{0}
 \definecolor{WHITE}{gray}{1}
 \definecolor{RED}{rgb}{1,0,0}
 \definecolor{GREEN}{rgb}{0,1,0}
 \definecolor{BLUE}{rgb}{0,0,1}
 \definecolor{CYAN}{cmyk}{1,0,0,0}
 \definecolor{MAGENTA}{cmyk}{0,1,0,0}
 \definecolor{YELLOW}{cmyk}{0,0,1,0}
}

\makeatother

\usepackage{babel}
\begin{document}

\title{Andreev reflection assisted lasing in an electromagnetic resonator
coupled to a hybrid-quantum-dot }

\author{S. Mojtaba Tabatabaei}

\email{s.m.taba90@gmail.com}

\selectlanguage{english}%

\affiliation{Department of Physics, Faculty of Sciences, Shahid Beheshti University,
G. C. Evin, Tehran 1983963113, Iran}

\author{Farshad Ebrahimi}

\affiliation{Department of Physics, Faculty of Sciences, Shahid Beheshti University,
G. C. Evin, Tehran 1983963113, Iran}
\begin{abstract}
A single mode electromagnetic resonator coupled to a two-level hybrid-quantum-dot(hQD)
is studied theoretically as a laser(maser), when the hQD is driven
out of equilibrium with external applied d.c. bias voltage. Using
the formalism of the non-equilibrium Green's functions for the hQD
and the semi-classical laser equations, we determine the relevant
physical quantities of the system. We find that due to the resonant
Andreev reflections and the formation of the Floquet-Andreev side-resonances
in the sub-gap region, at appropriate gate voltages and above a certain
threshold bias voltage and damping factor of the resonator, the two-level
QD has non-zero gain spectrum and lasing can happen in the system
in the frequency range of superconducting gap. Furthermore, our results
show that depending on the damping factor of the resonator and above
a specific threshold bias voltages, the lasing can be either due to
single electron transitions or cascaded electron transitions between
the Andreev resonances and Floquet-Andreev side-resonances.
\end{abstract}
\maketitle

\section{Introduction}

Recent developments in the nanotechnologies have made it feasible
to fabricate QDs coupled to a microwave resonator on a chip\cite{delbecq2011coupling,PhysRevX.1.021009,frey2012dipole,PhysRevB.85.045446,PhysRevB.89.195127}.
Among many theoretical and experimental aspects of the interaction
of electromagnetic waves of resonator with QD which have been studied,
the possibility of creating lasing in an electromagnetic resonator
using QDs has attracted considerable interest. 

Different proposals for achieving lasing in electromagnetic resonators
coupled to QDs have been considered. Jin et. al\cite{jin2011lasing},
liu et. al\cite{liu2015semiconductor} and Karlewski et.al\cite{karlewski2016lasing}
have shown that in a double-QD connected to metallic leads at finite
bias, population inversion can be created by electron tunneling. In
Ref.\onlinecite{marthaler2011lasing}, Marthaler and his coworkers
have shown that lasing without inversion can be achieved by coupling
the system to a dissipative environment which enhances the photon
emission. Lasing without inversion by coherently driving the system,
has recently being considered in Ref.\onlinecite{yuan2014sideband}
for a three-level V-type QD connected to external leads at finite
bias. Also, lasing was reported in Ref.\onlinecite{stehlik2016double}
by coupling the electrons of QD to external periodic driving field.
The periodic external field generates Floquet ladder which consist
of a series of doublet side-band of dressed-states. The inversion-less
gain spectrum in such a system is due to the unequalness of relative
populations of doublet dressed-states. Bruhat et.al\cite{bruhat2016cavity}
have also reported optical gain at finite bias in a single-level hybrid-QD\cite{sun1999resonant,PhysRevLett.87.176601,PhysRevB.63.094515,krawiec2003electron,bai2011andreev,zhang2012phonon,allub2015hybrid,weymann2015andreev,hwang2016hybrid,nussbaumer2004quantum,deacon2010tunneling,PhysRevB.81.121308,dirks2011transport}
which consists of a QD coupled to a normal metal and a superconducting
electrode. They showed that if the coupling of the QD to the superconducting
electrode is weak enough to suppress the Andreev reflections and widening
the width of the density of states at the two edges of the superconducting
gap, optical gain can be achieved.

In this work, we consider a single mode electromagnetic resonator
coupled to a two-level hQD where the coupling between the superconducting
electrode and the QD is not weak. The new features arising in this
hybrid system, due to interplay of the fundamental electronic interactions
and the proximity effects, are the formation of resonant Andreev reflections
and their Flouquet side-resonances and the possibility of sub-gap
transport. We show that at appropriate gate voltages and the damping
factor of the resonator and above a threshold bias voltage, the Andreev
resonances and their Floquet side-resonances in the sub-gap have unequal
populations and lasing can be achieved in the frequency range of superconducting
gap. 

Using the formalism of non-equilibrium Green's functions at zero temperature,
we, at first, determine numerically the linear gain spectrum of the
two-level hQD as a function of frequency and gate voltage for a fixed
bias voltage. Then, by solving numerically the semi-classical laser
equations self-consistently, we determine the lasing regimes, the
time-averaged and time-dependent currents through hQD and the photon
populations in the resonator in terms of d.c applied bias and gate
voltages for two different configurations of the energy levels of
the QD and damping factors of the resonator.

This paper is organized as follows. In Sec.\ref{sec:The-model}, we
introduce our model Hamiltonian and derive the related non-equilibrium
Green's functions for a two-level hQD coupled to a single mode electromagnetic
resonator. In Sec.\ref{sec:Physical-quantities}, we give the necessary
relevant formulas for various physical quantities such as average
photon number, electron occupations, current through hQD and etc.
. Finally, we present our numerical results and conclusions in Sec.\ref{sec:Results-and-conclusions}.

\section{\label{sec:The-model}The model}

\begin{figure}
\includegraphics[width=8.6cm]{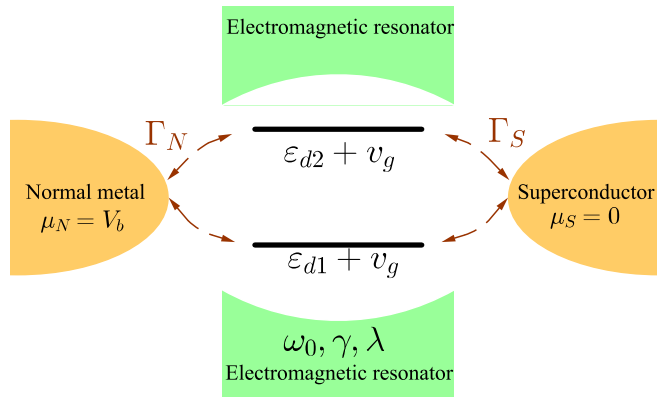}

\protect\caption{\label{figsys}(Color online) A single mode electromagnetic resonator
with frequency $\omega_{0}$, and damping factor $\gamma$, dipole
coupled with coupling constant $\lambda$, to a hQD consisting of
a two-level QD connected to a metallic and a superconducting electrodes.}
\end{figure}
Figure \ref{figsys} shows a schematic view of our model. We consider
a two-level quantum dot dipole coupled to the electric field of a
single mode electromagnetic resonator. The two levels of the QD, which
we assume to have different parities, are coupled to two electrodes,
a superconducting and a normal metal. Therefore, the total Hamiltonian
of our model, $H_{M}$, is described by the sum of the following terms:
\begin{alignat}{1}
H_{QD}= & \underset{n=1,2}{\sum}\,\underset{\sigma=\uparrow,\downarrow}{\sum}\left(\varepsilon_{d,n}+v_{g}\right)\, d_{n,\sigma}^{\dagger}d_{n,\sigma},\label{eq:1}\\
H_{leads}= & \underset{k,\sigma}{\sum}\left(\varepsilon_{k}+\mu_{N}\right)\, c_{k,\sigma}^{\dagger}c_{k,\sigma}\nonumber \\
 & +\underset{k,\sigma}{\sum}\left(\tilde{\varepsilon}_{k}+\mu_{S}\right)\, f_{k,\sigma}^{\dagger}f_{k,\sigma}\nonumber \\
 & +\underset{k}{\sum}\Delta\left(f_{k,\uparrow}^{\dagger}f_{-k,\downarrow}^{\dagger}+h.c.\right),\label{eq:2}\\
H_{T}= & \underset{k,n,\sigma}{\sum}t_{N}\,\left(c_{k,\sigma}^{\dagger}d_{n,\sigma}+h.c.\right)+t_{S}\,\left(f_{k,\sigma}^{\dagger}d_{n,\sigma}+h.c.\right),\label{eq:3}\\
H_{ph}=\hbar & \omega_{0}\left(a^{\dagger}a+\frac{1}{2}\right)\label{eq:4}
\end{alignat}
and 
\begin{gather}
H_{int}=-\underset{\sigma}{\sum}\lambda\left(a+a^{\dagger}\right)\left(d_{1,\sigma}^{\dagger}d_{2,\sigma}+h.c.\right),\label{eq:5}
\end{gather}
where, $H_{QD}$ is the Hamiltonian of isolated two-level QD, $H_{leads}$
is the sum of Hamiltonians of normal and superconducting leads, $H_{T}$
is the tunnelings Hamiltonian of the QD with the electrodes, $H_{ph}$
is the Hamiltonian of single mode electromagnetic resonator and $H_{int}$
is the interaction Hamiltonian of the electric field of resonator
with the electric dipole moment of the QD.

In Eqs.(\ref{eq:1}-\ref{eq:3}), $d_{n,\sigma}^{\dagger}(d_{n,\sigma}),$
$c_{k,\sigma}^{\dagger}(c_{k,\sigma})$ and $f_{k,\sigma}^{\dagger}(f_{k,\sigma})$
are, respectively, the fermionic creation(annihilation) operators
with spin $\sigma$ of QD, normal metal lead and superconducting lead,
$\varepsilon_{d,n}$, $n=1,2$, $\tilde{\varepsilon}_{k}$ and $\varepsilon_{k}$
are the orbital energies, $v_{g}=\tilde{v}_{g}+(\mu_{N}+\mu_{S})/2$,
$\tilde{v}_{g}$ is the external gate voltage applied to the QD, $\mu_{N}$
and $\mu_{S}$ are the chemical potentials of normal and superconducting
leads, $\Delta$ is the superconducting order parameter and $t_{N}$
and $t_{S}$ are the hybridization constants between the QD and the
normal and superconducting leads. In Eqs. (\ref{eq:4}) and (\ref{eq:5}),
$a^{\dagger}(a)$ is the photon creation(annihilation) operator, $\omega_{0}$
is the frequency of the resonator and $\lambda$ is the electric dipole
coupling strength of QD and the photon of the resonator.

We determine the possibility of lasing in the resonator of our model,
using the semi-classical laser equations\cite{haken2012laser}. The
Heisenberg equation of motion in the mean-field approximation for
the mean value of the annihilation operator of the photon is
\begin{equation}
i\hbar\frac{d}{dt}\left\langle a_{H}(t)\right\rangle =\hbar\omega_{0}\left\langle a_{H}(t)\right\rangle -\lambda\underset{\sigma}{\sum}\left\langle \left(d_{H1,\sigma}^{\dagger}(t)d_{H2,\sigma}(t)+h.c.\right)\right\rangle ,\label{eq:6}
\end{equation}
where all the operators are in the Heisenberg representation. The
semi-classical laser equations can be deduced from the above equation
by adding a phenomenological damping term, $-i\hbar\gamma\left\langle a_{H}(t)\right\rangle $,
to mimic the resonator's losses and separating the fast and slow parts
of the averaged quantities, using the slowly varying amplitude and
phase approximation\cite{haken2012laser}, where the mean values are
represented in the following forms
\begin{equation}
\left\langle a_{H}(t)\right\rangle =A_{\bar{\omega}}(t)e^{-i\phi(t)}e^{-i\bar{\omega}t}\label{eq:7}
\end{equation}
and
\begin{equation}
\underset{\sigma}{\sum}\left\langle \left(d_{H1,\sigma}^{\dagger}(t)d_{H2,\sigma}(t)+h.c.\right)\right\rangle =P_{\bar{\omega}}(t)e^{-i\phi(t)}e^{-i\bar{\omega}t}.\label{eq:8}
\end{equation}
In the above equations, $e^{-i\bar{\omega}t}$ is the fast oscillating
part, $A_{\bar{\omega}}(t)e^{-i\phi(t)}$ and $P_{\bar{\omega}}(t)e^{-i\phi(t)}$
are the slowly varying parts. Substituting expressions (\ref{eq:7})
and (\ref{eq:8}) int Eq. (\ref{eq:6}) and separating its real and
imaginary parts, we obtain
\begin{equation}
\frac{d}{dt}A_{\bar{\omega}}(t)=-\gamma A_{\bar{\omega}}(t)-\frac{\lambda}{\hbar}Im[P_{\bar{\omega}}(t)]\label{eq:9}
\end{equation}
and
\begin{equation}
\frac{d}{dt}\phi(t)=\hbar(\omega_{0}-\bar{\omega})-\lambda\frac{Re[P_{\bar{\omega}}(t)]}{A_{\bar{\omega}}(t)}.\label{eq:10}
\end{equation}
The above equations are the semi-classical laser equations. Their
stationary solutions, i.e $\frac{d}{dt}A_{\bar{\omega}}(t)$ and $\frac{d}{dt}\phi(t)$
equal to zero, which must be obtained self consistently with $P_{\bar{\omega}}$,
gives the laser threshold, the field intensity or the average photon
population and the frequency pulling of the resonator\cite{haken2012laser}.

To determine the steady-state solutions of Eqs. (\ref{eq:9}) and
(\ref{eq:10}), we compute the $P_{\bar{\omega}}$ and the other relevant
physical quantities of hQD by employing the non-equilibrium Green's
functions method. The usage of the non-equilibrium Green's functions
method allows us to take into account the effect of the electrodes
on the QD to infinite order of tunneling processes between the QD
and the electrodes. This offers an advantage over the conventional
quantum master equation method\cite{ginzel1993quantum,konig1996zero},
in which the coupling of the electrodes with the QD is treated to
the first order processes (weak coupling) or at most the next-to-the-leading
order tunneling processes. Furthermore, we use the exact form of the
interaction Hamiltonian of the electric field of the resonator with
the electric-dipole moment of the QD, which is more accurate and convenient
for numerical calculations than the usual dipole Hamiltonian in the
rotating wave approximation.

Within the mean-field approximation for the electric field in the
resonator, the interaction part of Hamiltonian reduces to
\begin{gather}
\tilde{H}_{int}(t)=\underset{\sigma}{-\lambda\sum}2A_{\bar{\omega}}cos(\bar{\omega}t)\left(d_{1,\sigma}^{\dagger}d_{2,\sigma}+h.c.\right).
\end{gather}

We study the case that the superconducting lead is grounded and an
static external bias voltage of $V_{b}$ is applied to the normal
lead. Furthermore, we work in units where $\hbar=e=c=1$. It is evident
that the explicit time dependence of the total Hamiltonian is only
through $\tilde{H}_{int}\left(t\right)$ which has a harmonic time
dependence with period $\frac{2\pi}{\bar{\omega}}$. So, it is convenient
to use Floquet representation
\begin{equation}
\mathcal{F}\left(t,t'\right)=\stackrel[m,n=-\infty]{+\infty}{\sum}\int_{-\frac{\bar{\omega}}{2}}^{\frac{\bar{\omega}}{2}}\frac{d\omega}{2\pi}e^{-i\left(\omega+m\bar{\omega}\right)t}e^{i\left(\omega+n\bar{\omega}\right)t'}\mathcal{F}_{mn}\left(\omega\right),
\end{equation}
for calculating different Green's functions and self-energies of the
system.

Using Nambu representation, $\Psi^{\dagger}=\left(d_{1,\uparrow}^{\dagger},d_{1,\downarrow},d_{2,\uparrow}^{\dagger},d_{2,\downarrow}\right)$,
the Fourier transform of the non-interacting retarded Green's function,
$\left[g^{R}\left(t,t'\right)\right]\equiv-i\theta\left(t-t'\right)\left\langle \left\{ \Psi\left(t\right),\Psi^{\dagger}\left(t'\right)\right\} \right\rangle _{0}$,
is given by\cite{sun1999resonant,sun1999photon}
\begin{equation}
\left[g_{mn}^{R}\left(\omega\right)\right]=\delta_{mn}\left[\left(\omega_{m}+i\eta\right)\left[I\right]-\left[h_{d}\right]-\left[\Sigma_{mn}^{R}\left(\omega_{m}\right)\right]\right]^{-1},
\end{equation}
where $\delta_{mn}$ is the Kronecker delta, $\omega_{m}=\omega+m\bar{\omega}$,
$\eta$ is an infinitesimal positive constant and $\left[h_{d}\right]$
is a $4\times4$ diagonal matrix with diagonal elements $\left(\varepsilon_{d,1}+v_{g},-\varepsilon_{d,1}-v_{g},\varepsilon_{d,2}+v_{g},-\varepsilon_{d,2}-v_{g}\right)$.
In the sequel the quantities in the brackets represent $4\times4$
matrices in the Nambu space. Furthermore, the effect of two electrodes
on QD is expressed by the self-energies of leads, $\left[\Sigma_{mn}^{R}\left(\omega\right)\right]=\left[\Sigma_{S,mn}^{R}\left(\omega\right)\right]+\left[\Sigma_{N,mn}^{R}\left(\omega\right)\right]$
which are\cite{trocha2014spin}
\begin{equation}
\left[\Sigma_{S,mn}^{R}\left(\omega\right)\right]=\delta_{mn}\left(\begin{array}{cccc}
a & b & a & b\\
b & a & b & a\\
a & b & a & b\\
b & a & b & a
\end{array}\right)\label{eq:self_ret_s}
\end{equation}
and
\begin{equation}
\left[\Sigma_{N,mn}^{R}\left(\omega\right)\right]=-\delta_{mn}i\Gamma_{N}\left(\begin{array}{cccc}
1 & 0 & 1 & 0\\
0 & 1 & 0 & 1\\
1 & 0 & 1 & 0\\
0 & 1 & 0 & 1
\end{array}\right),\label{eq:self_ret_n}
\end{equation}
where $a=-i\Gamma_{S}\beta\left(\omega\right)$ and $b=i\Gamma_{S}\beta\left(\omega\right)\frac{\Delta}{\omega}$.
The parameter $\beta\left(\omega\right)$ which is related to the
normalized BCS density of states is given by $\beta\left(\omega\right)=\frac{\left|\omega\right|}{\sqrt{\omega^{2}-\Delta^{2}}}\theta\left(\left|\omega\right|-\Delta\right)-i\frac{\omega}{\sqrt{\Delta^{2}-\omega^{2}}}\theta\left(\Delta-\left|\omega\right|\right)$.
We use the wide-band approximation where the hybridization of QD orbitals
with electrodes take the simple form $\Gamma_{N,S}\equiv\pi|t_{N,S}|^{2}\rho_{0}^{N,S}$
where $\rho_{0}^{N}$ and $\rho_{0}^{S}$ are the frequency independent
density of states of the normal lead and the normal state of the SC
lead, respectively.

We use the Dyson equation in the Floquet basis
\begin{equation}
\left[G_{mn}^{R}\left(\omega\right)\right]=\left[g_{mn}^{R}\left(\omega\right)\right]+\underset{lr}{\sum}\left[g_{ml}^{R}\left(\omega\right)\right]\left[\Pi_{lr}^{R}\left(\omega\right)\right]\left[G_{rn}^{R}\left(\omega\right)\right],
\end{equation}
to obtain the interacting retarded Green's function, $\left[G_{mn}^{R}\left(\omega\right)\right]$,
of the QD. The $\left[\Pi_{lr}^{R}\left(\omega\right)\right]$ in
the Dyson equation is the retarded self-energy, due to the interaction
term of the Hamiltonian, which has the form 
\begin{equation}
\left[\Pi_{lr}^{R}\left(\omega\right)\right]=-\lambda A\left(\delta_{l,r+1}+\delta_{l,r-1}\right)\left(\begin{array}{cccc}
0 & 0 & 1 & 0\\
0 & 0 & 0 & -1\\
1 & 0 & 0 & 0\\
0 & -1 & 0 & 0
\end{array}\right).
\end{equation}

Next, we need to calculate the lesser Green's function $G^{<}\left(t,t'\right)\equiv i\left\langle \Psi^{\dagger}\left(t'\right)\Psi\left(t\right)\right\rangle $.
We use Keldysh relation for lesser Green's function which in the Floquet
basis is
\begin{equation}
\left[G_{mn}^{<}\left(\omega\right)\right]=\underset{lr}{\sum}\left[G_{ml}^{R}\left(\omega\right)\right]\left[\Sigma_{lr}^{<}\left(\omega\right)\right]\left[G_{rn}^{A}\left(\omega\right)\right].
\end{equation}
Here, $\left[G_{rn}^{A}\left(\omega\right)\right]$ is the advanced
Green's function given by $\left[G_{rn}^{A}\left(\omega\right)\right]=\left[G_{rn}^{R}\left(\omega\right)\right]^{\dagger}$,
and $\left[\Sigma_{mn}^{<}\left(\omega\right)\right]=\left[\Sigma_{S,mn}^{<}\left(\omega\right)\right]+\left[\Sigma_{N,mn}^{<}\left(\omega\right)\right]$
is the lesser self-energy due to the coupling of QD to the electrodes
where 
\begin{equation}
\left[\Sigma_{S,mn}^{<}\left(\omega\right)\right]=\left[\left[\Sigma_{S,mn}^{A}\left(\omega\right)\right]-\left[\Sigma_{S,mn}^{R}\left(\omega\right)\right]\right]f\left(\omega\right)
\end{equation}
 and
\begin{equation}
\left[\Sigma_{N,mn}^{<}\left(\omega\right)\right]=\delta_{mn}2i\Gamma_{N}\left(\begin{array}{cccc}
f^{+}\left(\omega\right) & 0 & f^{+}\left(\omega\right) & 0\\
0 & f^{-}\left(\omega\right) & 0 & f^{-}\left(\omega\right)\\
f^{+}\left(\omega\right) & 0 & f^{+}\left(\omega\right) & 0\\
0 & f^{-}\left(\omega\right) & 0 & f^{-}\left(\omega\right)
\end{array}\right),
\end{equation}
with $f\left(\omega\right)=\theta\left(-\omega\right)$ and $f^{\pm}\left(\omega\right)=\theta\left(V_{b}\mp\omega\right)$,
which are the Fermi-Dirac distribution functions for the superconducting
and normal metal leads at zero temperature.

\section{\label{sec:Physical-quantities}Physical quantities}

We now, present the various relevant physical quantities related to
our model system using different Green's functions. The first quantities
of interest are the average polarization $\left\langle P\left(t\right)\right\rangle =\underset{\sigma}{\sum}\left\langle \left(d_{H1,\sigma}^{\dagger}(t)d_{H2,\sigma}(t)+h.c.\right)\right\rangle $
which is related to the lesser Green's function and the linear optical
susceptibility
\begin{equation}
\chi_{e}^{r}\left(t-t'\right)=-i\theta\left(t-t'\right)\left\langle \left[P(t'),P(t)\right]\right\rangle _{o},
\end{equation}
where $\left\langle ...\right\rangle _{o}$ indicates expectation-value
with respect to the non-interacting ground-state of the hQD. Using
the definition of the lesser Green's function, we get 
\begin{gather}
\left\langle P\left(t\right)\right\rangle =-i\left[G^{<}(t,t)\right]_{13+31-24-42}
\end{gather}
or
\begin{equation}
\left\langle P\left(t\right)\right\rangle =-i\underset{m,n}{\sum}\int_{-\frac{\bar{\omega}}{2}}^{\frac{\bar{\omega}}{2}}\frac{d\omega}{2\pi}e^{-i(m-n)\bar{\omega}t}\left[G_{mn}^{<}\left(\omega\right)\right]_{13+31-24-42},
\end{equation}
where the subscripts outside brackets represent different matrix elements
in the Nambu space which must be summed up. Setting the constant phase,
$\phi$, to zero and using Eq.(\ref{eq:8}), we obtain for the steady-state
amplitude of the polarization and the linear optical susceptibility
\begin{equation}
P_{\bar{\omega}}=-i\underset{m}{\sum}\int_{-\frac{\bar{\omega}}{2}}^{\frac{\bar{\omega}}{2}}\frac{d\omega}{2\pi}\left[G_{m+1,m}^{<}\left(\omega\right)\right]_{13+31-24-42},\label{eq:24}
\end{equation}
and\cite{haug2008quantum} 
\begin{align}
\chi_{e}^{r}\left(\omega\right)=( & \boldsymbol{F}_{33}^{11}\left(\omega\right)+\boldsymbol{F}_{13}^{13}\left(\omega\right)-\boldsymbol{F}_{23}^{14}\left(\omega\right)-\boldsymbol{F}_{43}^{12}\left(\omega\right)+\nonumber \\
 & \boldsymbol{F}_{11}^{33}\left(\omega\right)+\boldsymbol{F}_{31}^{31}\left(\omega\right)-\boldsymbol{F}_{41}^{32}\left(\omega\right)-\boldsymbol{F}_{21}^{34}\left(\omega\right)+\nonumber \\
 & \boldsymbol{F}_{44}^{22}\left(\omega\right)+\boldsymbol{F}_{24}^{24}\left(\omega\right)-\boldsymbol{F}_{34}^{21}\left(\omega\right)-\boldsymbol{F}_{14}^{23}\left(\omega\right)+\nonumber \\
 & \boldsymbol{F}_{22}^{44}\left(\omega\right)+\boldsymbol{F}_{42}^{42}\left(\omega\right)-\boldsymbol{F}_{12}^{43}\left(\omega\right)-\boldsymbol{F}_{32}^{41}\left(\omega\right)),\label{eq:25}
\end{align}
where 
\begin{equation}
\boldsymbol{F}_{pq}^{mn}\left(\omega\right)\equiv-i\underset{l}{\sum}\int\frac{d\omega'}{2\pi}\{\left[G_{0l}^{R}\left(\omega+\omega'\right)\right]_{mn}\left[G_{l0}^{<}\left(\omega'\right)\right]_{pq}+\left[G_{0l}^{<}\left(\omega+\omega'\right)\right]_{mn}\left[G_{l0}^{A}\left(\omega'\right)\right]_{pq}\}.
\end{equation}
Furthermore, the time-averaged expectation-value of the electron and
hole occupations of each orbital could be calculated using
\begin{equation}
\left\langle n_{d,m}\left(t\right)\right\rangle _{t}=-2i\underset{l}{\sum}\int_{-\frac{\bar{\omega}}{2}}^{\frac{\bar{\omega}}{2}}\frac{d\omega}{2\pi}\left[G_{ll}^{<}\left(\omega\right)\right]_{mm}\left(\omega\right),\: m=1,..,4
\end{equation}
where $\left\langle ...\right\rangle _{t}$ means time-averaged expectation
value, $m=1,2$ and $m=3,4$ designate, respectively, the electron
and hole states of the first and second levels of QD and the factor
two is due to the electron's spin. Moreover, the time-averaged total
density of states(DOS) of the QD could be obtained from retarded Green's
function as
\begin{equation}
\rho\left(\omega\right)=-\frac{1}{\pi}Tr\left[Im\left[G_{00}^{R}\left(\omega\right)\right]\right].\label{eq:dos}
\end{equation}
where $Tr\left[...\right]$ represents trace with respect to the Nambu
indeices. Finally, for calculating the time-dependent and time-averaged
electric current through the QD in terms of the Green's functions
and the self-energies in Floquet representations, we use the following
expressions\cite{sun1999resonant,wang2006conservation}; 
\begin{align}
I(t)=\underset{l,m,n}{\sum}\int_{-\frac{\bar{\omega}}{2}}^{\frac{\bar{\omega}}{2}}\frac{d\omega}{2\pi}e^{-i(l-n)\bar{\omega}t}[ & \left[G_{lm}^{R}\left(\omega\right)\right]\left[\Sigma_{N,mn}^{<}\left(\omega\right)\right]+\left[G_{lm}^{<}\left(\omega\right)\right]\left[\Sigma_{N,mn}^{A}\left(\omega\right)\right]\nonumber \\
 & -\left[\Sigma_{N,lm}^{<}\left(\omega\right)\right]\left[G_{mn}^{A}\left(\omega\right)\right]-\left[\Sigma_{N,lm}^{R}\left(\omega\right)\right]\left[G_{mn}^{<}\left(\omega\right)\right]]_{11-22+33-44}.
\end{align}
and
\begin{align}
\left\langle I\left(t\right)\right\rangle _{t}=\underset{l,m}{\sum}\int_{-\frac{\bar{\omega}}{2}}^{\frac{\bar{\omega}}{2}}\frac{d\omega}{2\pi}[ & \left[G_{lm}^{R}\left(\omega\right)\right]\left[\Sigma_{N,ml}^{<}\left(\omega\right)\right]+\left[G_{lm}^{<}\left(\omega\right)\right]\left[\Sigma_{N,ml}^{A}\left(\omega\right)\right]\nonumber \\
 & -\left[\Sigma_{N,lm}^{<}\left(\omega\right)\right]\left[G_{ml}^{A}\left(\omega\right)\right]-\left[\Sigma_{N,lm}^{R}\left(\omega\right)\right]\left[G_{ml}^{<}\left(\omega\right)\right]]_{11-22+33-44}.
\end{align}

\section{\label{sec:Results-and-conclusions}Results and conclusions}

\begin{figure}
\includegraphics[width=8.6cm]{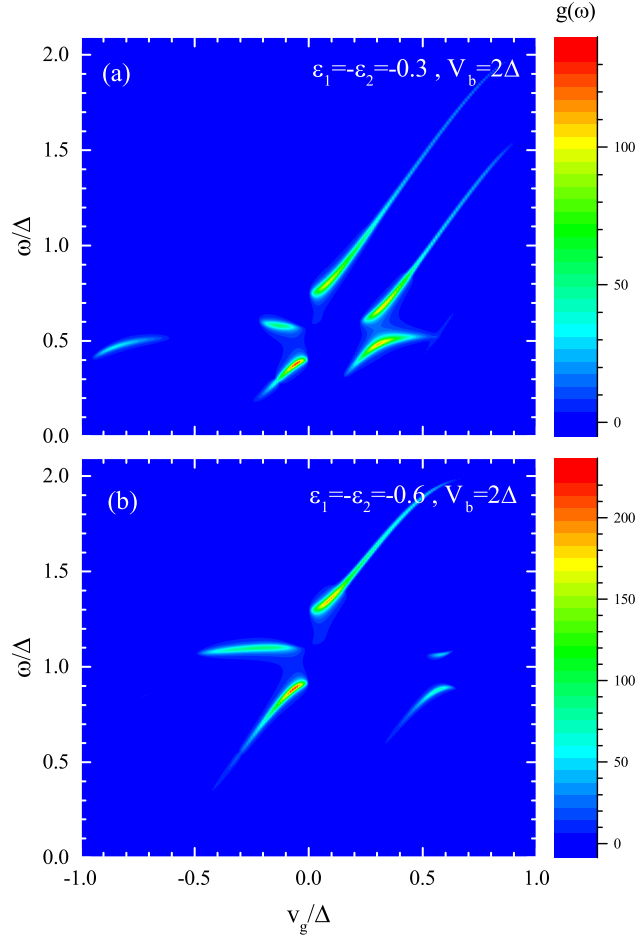}

\protect\caption{\label{figsys-1}(Color online) Linear gain spectrum of the QD as
functions of frequency and external gate voltage for (a) $\varepsilon_{d,1}=-\varepsilon_{d,2}=-0.3\Delta$,
(b) $\varepsilon_{d,1}=-\varepsilon_{d,2}=-0.6\Delta$, when the hQD
is externally biased at $V_{b}=2\Delta$. Other parameters are $\mu_{S}=0$,
$\mu_{N}=V_{b}=2\Delta$, $\Gamma_{S}=0.1\Delta$ and $\Gamma_{N}=0.01\Delta$.}
\end{figure}
In the preceding sections, the necessary formulas for determining
the lasing conditions for the hQD-resonator system were presented.
We now investigate the prospect of lasing in such a system. We start
by calculating, at first, the linear gain spectra, $g(\omega)=-4\pi\omega Im\left[\chi_{e}^{r}\left(\omega\right)\right]$,
of the QD which can be obtained from the imaginary part of the linear
optical susceptibility, given by Eq.(\ref{eq:25}). We consider the
following two different energy configurations for the two levels of
QD; $\varepsilon_{d,1}=-\varepsilon_{d,2}=-0.3\Delta$ and $\varepsilon_{d,1}=-\varepsilon_{d,2}=-0.6\Delta$.
The results as functions of $\omega/\Delta$ and $v_{g}$ for $\mu_{S}$
equal to zero, $\mu_{N}=V_{b}=2\Delta$, $\Gamma_{S}=0.1\Delta$ and
$\Gamma_{N}=0.01\Delta$, are depicted in Figs.\ref{figsys-1} (a)
and (b). Although one might expect to see non-zero gain only at frequencies
equal to the energy difference of the two levels of the QD but, as
we see in Fig.\ref{figsys-1}, this does not happen in our model system.
Instead, we see different regions for non-zero gain which are dependent
on the parameters of the QD. The origin of these gain regions is due
to the fact that it is, essentially, the electron transitions between
different resonant Andreev reflections in the sub-gap regions which
are responsible for the non-zero gain in the system. In Figs.\ref{figsys-1}
(a) and (b), in the first case, the maximum gain occurs at frequency
$\omega=0.38\Delta$ and gate voltage $v_{g}=-0.06\Delta$ and in
the second case, the maximum gain is at $\omega=0.87\Delta$ and $v_{g}=-0.05\Delta$.

\begin{figure}
\includegraphics[width=8.6cm]{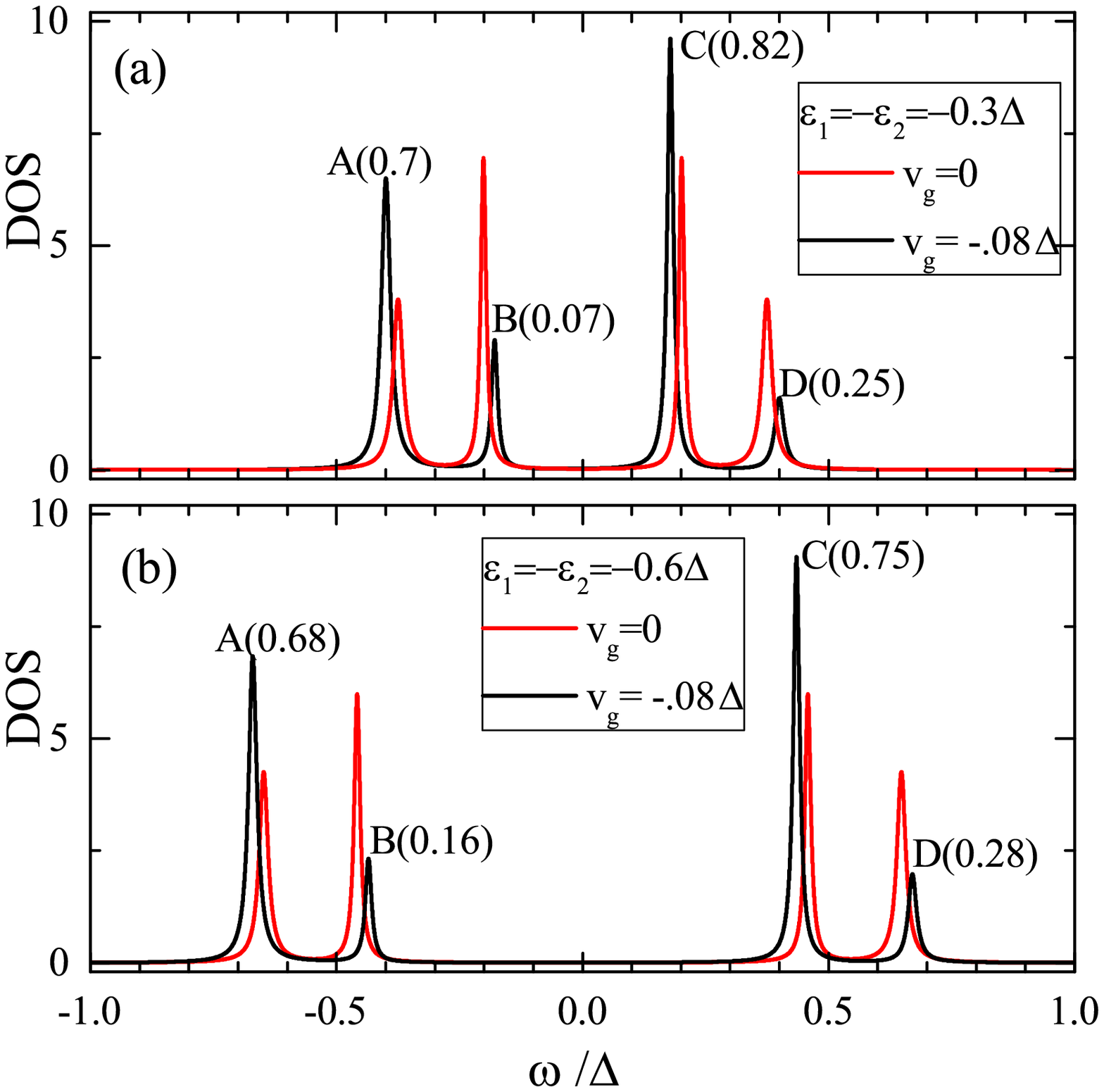}

\protect\caption{\label{figsys-1-1}(Color online) Total density of states of the QD
for (a) $\varepsilon_{d,1}=-\varepsilon_{d,2}=-0.3\Delta$, (b) $\varepsilon_{d,1}=-\varepsilon_{d,2}=-0.6\Delta$
at $v_{g}=-0.08\Delta$(solid-black line) where the maximum gain is
obtained and at $v_{g}=0$(red line). Other parameters are as in Fig.\ref{figsys-1}.
The numbers above each solid-black line peaks show their corresponding
electron population probabilities. The corresponding electron population
probabilities for $v_{g}=0$ peaks are $0.5$ .}
\end{figure}
In order to clarify the above discussion about the origin of the non-zero
gain in the system, we present in Figs.\ref{figsys-1-1} (a) and (b),
the density of states of QD and their relative populations for the
two aforementioned cases and compare them with the situations when
the gate voltages are zero. The four resonances in the density of
states are due to the Andreev reflections. It can be seen from Figs.
\ref{figsys-1-1} (a) and (b) that the relative populations of the
Andreev resonances are dependent on the applied external bias and
gate voltages and they could have some population inversions in the
sub-gap energies in some specific configurations. The maximum linear
gains in Fig.\ref{figsys-1}(a) and (b) are due to the transitions
from C to B resonances, depicted in Fig.\ref{figsys-1-1}(a) and (b),
respectively.

We next consider the possibility of lasing in a system of a single
mode electromagnetic resonator coupled to a hQD. We choose the aforementioned
configurations for hQD and two different damping factors; $\gamma=10^{-3}\Delta$
and $\gamma=10^{-4}\Delta$ for the resonator. We solve the semi-classical
laser Eqs.(\ref{eq:9}) and (\ref{eq:10}) with Eq.(\ref{eq:24})
for the polarization of hQD numerically and self-consistently.

\begin{figure}
\includegraphics[width=8.6cm]{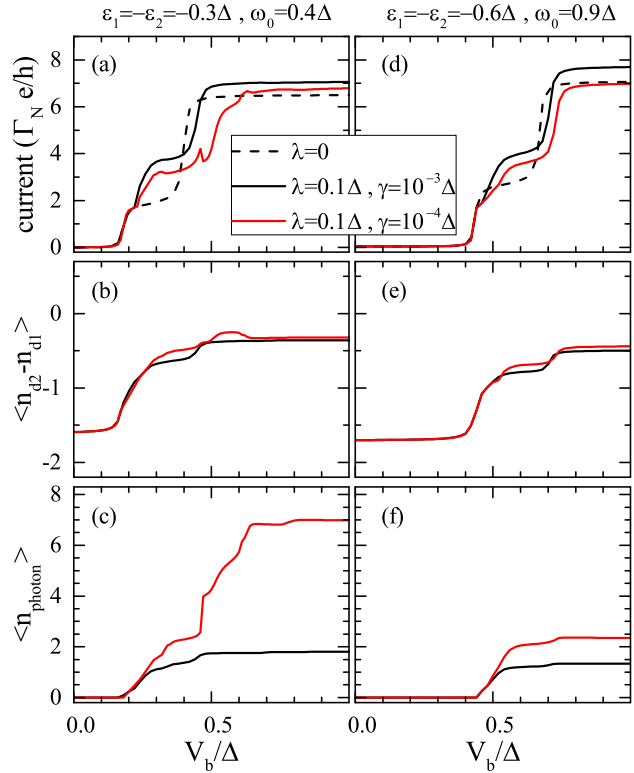}

\protect\caption{\label{figsys-1-1-1}(Color online) The time-averaged current, (a)
and (d), average population difference of the QD's orbitals, (b) and
(e), and average number of photons in the resonator, (c) and (f),
as functions of bias voltage for two different damping factors of
the resonator: $\gamma=10^{-3}\Delta$(black-solid line) and $\gamma=10^{-4}\Delta$(red-solid
line). The QD's orbitals and the bare frequency of the resonator are:
(left panels)$\varepsilon_{d,1}=-\varepsilon_{d,2}=-0.3\Delta,\omega_{0}=0.4\Delta$
and (right panels)$\varepsilon_{d,1}=-\varepsilon_{d,2}=-0.6\Delta,\omega_{0}=0.9\Delta$.
Dashed lines in (a) and (d) show the current in the absence of the
resonator. Other parameters are as in Fig.\ref{figsys-1}.}
\end{figure}
\begin{figure}
\includegraphics[width=8.6cm]{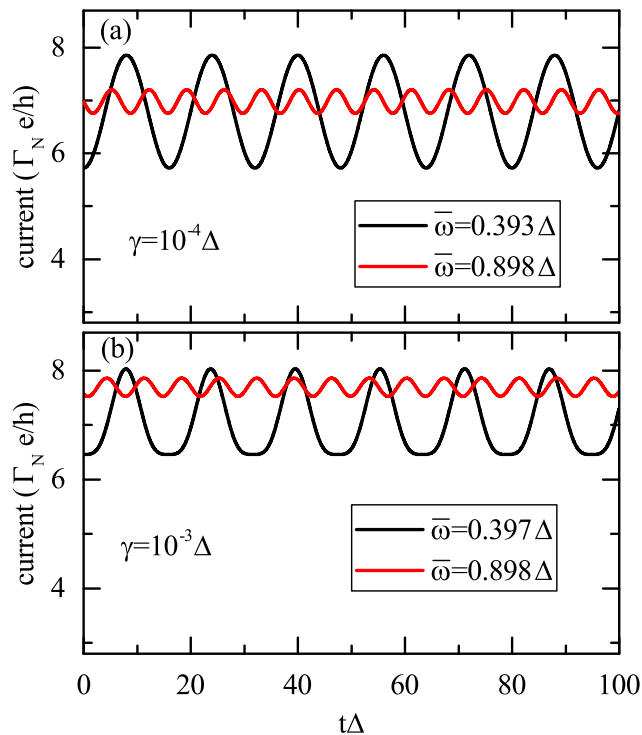}

\protect\caption{\label{figsys-1-1-1-1}(Color online) The time-dependent current through
QD in the lasing regime at $V_{b}=\Delta$ and $v_{g}=-0.08\Delta$
for $\varepsilon_{d,1}=-\varepsilon_{d,2}=-0.3\Delta$(black-solid
line) and $\varepsilon_{d,1}=-\varepsilon_{d,2}=-0.6\Delta$(red-solid
line) for two resonator damping factors; (a)$\gamma=10^{-4}\Delta$
and (b) $\gamma=10^{-3}\Delta$ . Other parameters are as in Fig.\ref{figsys-1}.}
\end{figure}

In Fig.\ref{figsys-1-1-1}, we have depicted the time-averaged current
through the QD, the average population differences of the two levels
of QD, and the average photon populations in the resonator as functions
of external applied bias for coupling constant $\lambda=0.1\Delta$.
In Figs.\ref{figsys-1-1-1}(a)-(c), the frequency of the resonator
is $\omega_{0}=0.4\Delta$ and in Figs\ref{figsys-1-1-1}(d)-(e),
$\omega_{0}=0.9\Delta$. Depending on the ratio of the intensity of
electric field in the resonator to the frequency of resonator and
the magnitude of applied bias voltage, we observe two regimes of lasing
for both cases. For $\gamma=10^{-3}\Delta$, we obtain small values
of the aforementioned ratio and the stimulated emission is solely
between the Andreev resonances and lasing occurs above a threshold
bias voltage. When we reduce the damping factor of the resonator to
$\gamma=10^{-4}\Delta$, the ratio of the intensity of the electric
field in the resonator to the frequency of the resonator becomes large
and the Floquet-Andreev side-resonances, with frequencies obeying
relation; $\omega_{m}=\omega_{A}+m\bar{\omega},m=0,\pm1,...,$ where
$\omega_{A}$'s are the frequencies of Andreev resonances, acquire
sizable amplitudes in the superconducting gap, and above certain threshold
bias voltage, their populations and frequency differences are such
that they can participate in the stimulated emission in two different
ways; either in a cascaded manner which results in a sudden increase
of the average number of photons in the resonator without appreciable
change in the time-averaged current through the QD, see Figs.\ref{figsys-1-1-1}(a)
and (c), or through extra electron transitions between the Floquet-Andreev
side-resonances which we observe in Figs\ref{figsys-1-1-1}(d) and
(f). In the latter case, the increase in the average number of photons
in the resonator is accompanied with an increase in the time-averaged
current through the QD. Furthermore, the on-set of lasing in the resonator
is accompanied by the appearance of oscillating polarization current
through the hQD which is depicted in Fig.\ref{figsys-1-1-1-1}.

In conclusion, we numerically investigated the possibility of lasing
in a single mode electromagnetic resonator coupled to a two-level
hQD when driven out of equilibrium by applying external bias voltages.
It is found that at specific gate voltages and above certain threshold
bias voltages the two-level QD connected to a normal metal and a superconducting
electrodes has non-zero gain spectrum due to the resonant Andreev
reflections and when coupled to an electromagnetic resonator, for
damping factors of the resonator below certain thresholds, Andreev-Floquet
side-resonances also appear in the sub-gap regions and lasing can
happens in two different regimes. In addition, with the on-set of
lasing in the resonator, the current through hQD beside its d.c. (time-averaged)
component, acquires an oscillating part. Thus, by monitoring the d.c.
and a.c. components of the current through the hQD, the on-set of
lasing and its regime can, in principle, be identified. \bibliographystyle{unsrt}
\bibliography{ref_lasing}

\end{document}